\documentclass[12pt, onecolumn]{IEEEtran} 
\usepackage{subfigure, epsfig, color}
\usepackage{amsmath}
\usepackage{amsthm}
\usepackage{amssymb}
\usepackage{multirow}
\usepackage{booktabs}

\usepackage{mathtools}

\begin{document}

\title{Deep Learning-Based Channel Estimation for High-Dimensional Signals}
\author{{
    Eren Balevi and
    Jeffrey G. Andrews}\\
    \IEEEauthorblockA{
		Department of Electrical and Computer Engineering \\
    The University of Texas at Austin, TX 78712, USA\\
    Email: erenbalevi@utexas.edu, jandrews@ece.utexas.edu}	
\thanks{This work has been supported in part by Intel.}				
}

\maketitle 
\normalsize
\begin{abstract}
We propose a novel deep learning-based channel estimation technique for high-dimensional communication signals that does not require any training. Our method is broadly applicable to channel estimation for multicarrier signals with any number of antennas, and has low enough complexity to be used in a mobile station.  The proposed deep channel estimator can outperform LS estimation with nearly the same complexity, and approach MMSE estimation performance to within $1$ dB without knowing the second order statistics. The only complexity increase with respect to LS estimator lies in fitting the parameters of a deep neural network (DNN) periodically on the order of the channel coherence time. We empirically show that the main benefit of this method accrues from the ability of this specially designed DNN to exploit correlations in the time-frequency grid. The proposed estimator can also reduce the number of pilot tones needed in an OFDM time-frequency grid, e.g. in an LTE scenario by $98\%$ ($68\%$) when the channel coherence time interval is $73$ms ($4.5$ms). 
\end{abstract}

\begin{IEEEkeywords}
Deep learning, channel estimation, massive MIMO, OFDM.
\end{IEEEkeywords}

\section{Introduction}
Obtaining accurate channel state information (CSI) with rudimentary estimation methods and a reasonable number of pilots has been and continues to be of great importance in cellular communication systems. However, this is quite challenging, especially for high-dimensional signals in time, frequency and space. For example, achieving satisfactory performance with the low complexity least-squares (LS) estimator is often impossible, while minimum mean square error (MMSE) channel estimation is very complex and difficult to achieve in practice. Motivated by the recent successes of deep learning for solving hard detection and estimation problems \cite{BalAnd19}-\cite{Felix} and considering the significant practical benefits of a channel estimator that has (near) LS estimation complexity with (near) MMSE estimation performance, it seems quite intriguing to attempt the channel estimation problem via a deep neural network (DNN).

Conventional DNNs require a large number of parameters that have to be trained with large data sets \cite{Goodfellow}. This training data set corresponds to the pairs of received signal-pilot symbols for DNN-based channel estimation which implies that many pilots are needed for reliable channel estimation via DNNs. This is particularly true for high-dimensional communication signals such as MIMO-OFDM. Although the computational complexity of DNNs is quite reasonable given recent advances in specialized hardware, the pilot symbols consume over-the-air bandwidth and must be kept to a bare minimum. Thus, DNN's hunger for training data is the major obstacle to using them for channel estimation.

In this paper, we propose a novel DNN-based channel estimation technique that does not require any training. It is based upon the recently proposed deep image prior (DIP) model, which fits the parameters of a DNN on the fly instead of training beforehand  \cite{Ulyanov18}. This approach not only greatly reduces the training overhead, it also prevents the usual mismatch between training and testing phase. We adapt and tune this method to have near MMSE channel estimation performance with LS-type complexity for high-dimensional signals. More precisely, in the proposed method the specially designed DNN first generates a less noisy signal from the received signal, and then this generated signal is element-wise divided into the pilot symbols just like an LS estimator. It is worth emphasizing that pilots are only used while estimating the channel and not for training the DNN. Hence, there is no need for labeled data. The proposed channel estimator is quite different from other recent work that utilized off-the-shelf deep learning algorithms for channel estimation, whose efficiency are attributed to the large amount of training data \cite{HeLi18}, \cite{Wen18}. 

The main contribution of this paper therefore is to propose an efficient, scalable (in terms of DNN parameters) channel estimation technique for high-dimensional signals that can work with a small number of pilots.  It is applicable to SISO-OFDM and massive MIMO-OFDM, and everything in between. Our method outperforms the LS channel estimation and can approach the MMSE channel estimation performance without knowing the second-order statistics of the channel and noise. We empirically show that the main success behind this lies in exploiting the correlations among subcarriers, which are captured by the designed DNN and used as a priori information to reduce the noise in the signal. The proposed estimator is scalable, because the number of the DNN parameters for the proposed model is independent from the number of subcarriers, which is achieved by parameter sharing. Furthermore, our deep channel estimator yields very efficient pilot usage. As a specific example, the proposed estimator can reduce by $98\%$ ($68\%$) the number of pilot tons in an LTE time-frequency grid when the channel coherence time interval is $73$ms ($4.5$ms). 

The paper is organized as follows. The channel estimation problem is stated in Section \ref{Problem Statement}. The proposed deep channel estimator is presented in Section \ref{Deep Channel Estimator Model}, and its performance is thoroughly discussed in Section \ref{Simulations}. The paper ends with concluding remarks and future work in Section \ref{Conclusions}.  

\textit{Notation}: Scalars in the frequency domain are shown as uppercase plain letters throughout the paper. Vectors are designated as uppercase boldface letter. Transpose operation is demonstrated by $(\cdot)^T$. The matrix inverse is denoted as $(\cdot)^{-1}$. The set of real and complex numbers are represented by $\mathbb{R}$ and $\mathbb{C}$.

\section{Problem Statement} \label{Problem Statement}
The frequency domain representation of an OFDM signal between the $m^{th}$ receive and the $q^{th}$ transmit antenna for the $n^{th}$ OFDM symbol can be expressed as
\begin{equation}\label{OFDM_freq}
\mathbf{Y[m,n]} = \mathbf{H[m,q]}\circ \mathbf{X[q,n]}+ \mathbf{W[m,n]}
\end{equation}
where $\circ$ is the Hadamard product, $\mathbf{Y[m,n]} \in \mathbb{C}^{N_f}$, $\mathbf{H[m,q]} \in \mathbb{C}^{N_f}$ and $\mathbf{X[q,n]}  \in \mathbb{C}^{N_f}$ are the received signal, channel and transmitted signal for $N_f$ subcarriers. The time index of the channel is omitted for simplicity, since the channel remains constant for some number of OFDM symbols related with the coherence time. The last term $\mathbf{W[m,n]} \in \mathbb{C}^{N_f}$ represents the zero-mean Gaussian noise with variance $\sigma^2$. This expression can be easily generalized for interference by treating the $\mathbf{W[m,n]}$ as denoting the sum of Gaussian noise and interference that have some probability distribution. Additionally,
\begin{equation}
\mathbf{Y[m,n]}=[Y[1,m,n], \cdots, Y[N_f,m, n]]^{T}
\end{equation}
where $Y[k,n,m]$ is the $k^{th}$ subcarrier in the $\mathbf{Y[m,n]}$. This notation is the same for the channel and transmitted signal. In one coherence time interval, there are $N$ OFDM symbols in the time domain and $N_{f}$ number of subcarriers in the frequency domain. Thus, the OFDM time-frequency grid for $N_r$ receive antennas becomes
\begin{equation} \label{OFDM_grid}
\mathbf{Y} = \{\{\{Y[k,m,n]\}_{k=1}^{N_{f}}\}_{m=1}^{N_r}\}_{n=1}^{N}.
\end{equation}

Estimating the channel for OFDM was studied nearly two decades ago \cite{Beek95}, and it is very well-known that LS channel estimation has the lowest complexity at the expense of some performance loss. On the other hand, MMSE channel estimation has higher complexity, but  with much better performance. To make the complexity of the canonical LS and MMSE estimators concrete, these estimators are written in closed-form assuming that all the pilots are sent at the beginning of each coherence time interval. Using these pilots, LS channel estimation can be simply done by element-wise division of the first received OFDM symbol from the $m^{th}$ receive antenna with the first transmitted (pilot) OFDM symbol from the $q^{th}$ transmit antenna as
\begin{equation}\label{ls_chn_est}
\hat{H}_{LS}[k, m, q]= Y[k,m,1] / X[k,q,1] 
\end{equation}
for $k=\{1, \cdots, N_f\}$, $m=\{1, \cdots, N_r\}$ and $q=\{1, \cdots, N_t\}$. The LS estimator in \eqref{ls_chn_est} can be written as a vector of subcarriers for each transmit-receive antenna pair
\begin{equation}\label{LS_est_vect}
\mathbf{\hat{H}_{LS}[m,q]} = [\hat{H}_{LS}[1,m, q],  \cdots, \hat{H}_{LS}[N_f,m, q]]^{T}. 
\end{equation}

To quantify the complexity of MMSE channel estimation, it is computed as  
\begin{eqnarray} \label{mmse_chn_est}
\mathbf{\hat{H}_{MMSE}[m,q]} &=&\mathbf{R_{HH}[m,q]} \left(\rho\textbf{I}_{N_f} + \mathbf{R_{HH}[m,q]}\right)^{-1}\mathbf{\hat{H}_{LS}[m,q]} 
\end{eqnarray}
where $\rho$ is the inverse of the SNR, $\mathbf{R_{HH}[m,q]}$ is the auto-correlation matrix of the frequency domain channel taps between the  $m^{th}$ receive antenna and the $q^{th}$ transmit antenna, and $\textbf{I}_{N_f}$ is the $N_f \times N_f$ identity matrix. The matrix inversion and the need for the knowledge of the second-order statistics of the channel and noise render MMSE estimator impractical, especially for large number of antennas. For this reason, LS estimator is often only viable method for massive MIMO \cite{C.K.Wen15}. However, LS channel estimation yields less accurate estimation with respect to MMSE estimator, which can significantly decrease the performance. The impact of this difference in channel estimation becomes profound in massive MIMO \cite{NgoLarsMarz13}. Hence, novel low complexity high performance channel estimation methods are highly needed. 

Deep learning provides a new tool for channel estimation thanks to the various types of state-of-the-art neural network architectures. However, making these architectures operational for channel estimation is challenging because of the high dimensionality of communication signals. This is associated with the fact that the higher the signal dimension is, the larger the number of parameters in the DNN model must be, which needs to be trained with even larger data set proportional to the number of parameters. To illustrate, a fully connected neural network for a single antenna OFDM system requires $U=N_fN$ input neurons. If there are $l$ layers in this DNN, each of which has $k_iU$ units for $i=0, 1, \cdots, l-1$, this leads to $\sum_{i=0}^{l-1} k_ik_{i+1}U^2$ parameters, where $k_0=k_l=2$ due to the real and imaginary parts of the signal. This can yield millions of parameters, and thus requires a very large training data set. Although convolutional neural networks can considerably decrease the number of parameters, a large training data set is still necessary. This is obviously an impediment in using neural networks for channel estimation, which have to be done with very limited number of pilots (or labels)\footnote{There can be some unsupervised or semi-supervised learning models that make channel estimation with no labels or with very limited labels. However, there is not any known channel estimation model yet for this method, and this subject remains mostly open.}. 

In this paper, we propose a new DNN based channel estimation method in an attempt to approach the performance of MMSE estimator with nearly the same complexity of the LS estimator. The proposed method does not require any training, and thus it precludes the major obstacle to employing neural networks in channel estimation. It has also reasonable number of parameters, e.g., on the order of thousands or even hundreds, depending on the channel.  This relatively low complexity is largely due to the parameter sharing among subcarriers. This number of parameters is quite viable, even in a mobile handset, given recent advances towards efficient neural computing architectures. Additionally, our method prevents the performance loss due to the mismatch between training and testing phase in channel estimation, since there is no training. The details of the proposed method are elaborated upon next.

\section{Deep Channel Estimator Model} \label{Deep Channel Estimator Model} 
Training overhead is the primary obstacle to make state-of-the-art DNNs, which are mainly used for image and speech processing, operational (practically implementable) for high-dimensional channel estimation in practical receivers. Promisingly, a recent DNN model that was proposed to solve inverse problems in image processing efficiently does not require training \cite{Ulyanov18}. The main idea behind this model is to fit the parameters of a neural network for each image on the fly without training them on large data sets beforehand.  It is known as Deep Image Prior (DIP). This model was later optimized to reduce the number of required parameters \cite{Heckel18}. Both \cite{Ulyanov18} and \cite{Heckel18} observed very efficient denoising performance thanks to the specifically designed DNN architecture, which has low impedance for natural images and high impedance against noise. The general properties of this model, namely no need for training and denoising capability make it appealing for high-dimensional channel estimation in SISO-OFDM or in MIMO-OFDM due to the following reasons: (i) in communication systems, there is a limited number of pilots (or labeled data), and thus the architectures based upon large training data set are not feasible; (ii) in conventional DNNs, training and testing have to be done for the same channel realization to obtain better performance, which brings in heavy training overhead; and (iii) noise is one of the main impediments that hinders to make a reliable channel estimation. Motivated by these factors, the specifically designed DNN architecture for the DIP model is leveraged to make channel estimation. In particular, we modify the input and output layers of the one variant of DIP architecture \cite{Heckel18}, and use it as a baseline architecture, which is dubbed as \textit{deep channel estimator}.

The proposed deep channel estimator is based on the idea of generating a less noisy signal from the received signal through a specially designed DNN architecture, and then utilize this generated signal by the DNN for channel estimation. Specifically, this generated signal is divided to the pilot symbols in an element-wise fashion for channel estimation. This means that we propose an LS-type channel estimator with the only difference that the signal generated by the DNN is used instead of the received signal. By doing that the low complexity nature of LS estimator is combined with the noise reduction capability of the DNN so as to have a near MMSE estimation performance. The price paid for the proposed deep channel estimator is the need for fitting the parameters of the DNN periodically for each OFDM grid, which is defined as $\mathbf{Y}$ in \eqref{OFDM_grid}, whose period is determined by the channel coherence time (or equivalently maximum Doppler spread). However, the complexity increase is quite reasonable thanks to the low number of parameters that will be explained in detail. 

\begin{figure*} [!t] 
\centering 
\includegraphics [width=5in]{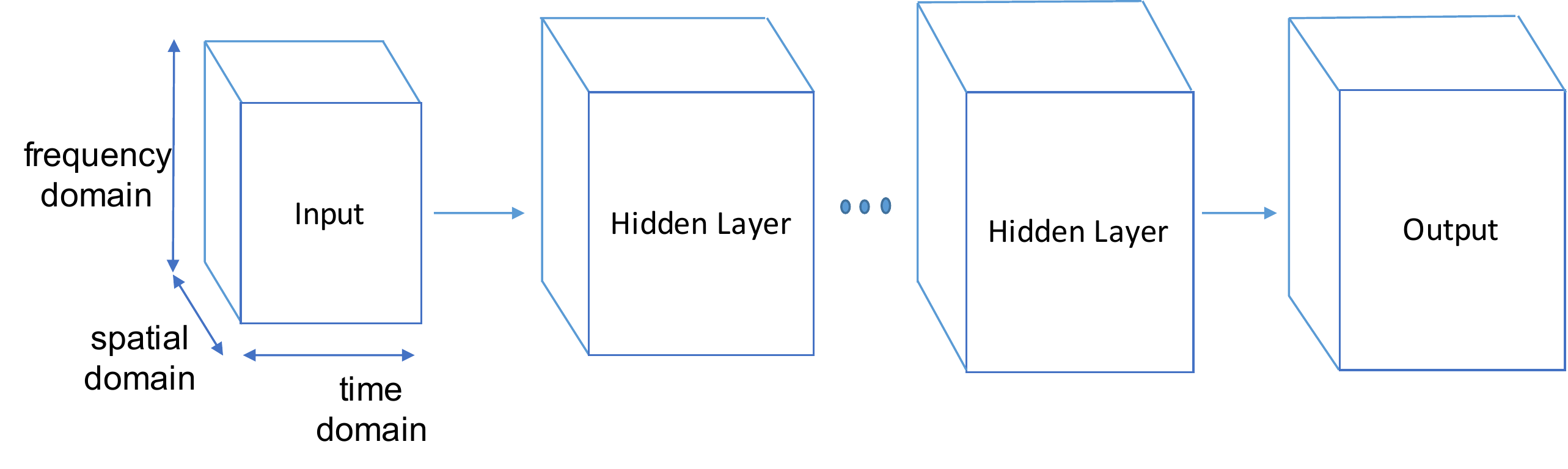}
\caption{The DNN architecture to denoise the received signal before channel estimation for a $3$-dimensional communication signal.} \label{fig:e2emodel}
\end{figure*}

The real and imaginary part of $\mathbf{Y}$ in \eqref{OFDM_grid} is separated into $2$ independent channels in our architecture, since tensors do not support complex operations. This tensor representation of $\mathbf{Y}$ is denoted as $\mathbf{Y_T}$. Specifically, $\mathbf{Y_T} \in \mathbb{R}^{N_f\times N \times N_r \times 2}$, where the dimensions are for the frequency, time, spatial and complex domains. Note that complex domain is to represent the real and imaginary parts of the signal (or equivalently in-phase and quadrature components). For the ease of notation $N_r \times 2$ is replaced by $N_s$, e.g., $\mathbf{Y_T} \in \mathbb{R}^{N_f\times N \times N_s}$. The main aim of the deep channel estimator is to generate the $\mathbf{Y_T}$ from a randomly chosen input tensor $Z_0$, which can be considered as an input filled with uniform noise, through hidden layers, whose weights are also randomly initialized, and then optimized via gradient descent. The overall DNN model that depicts the input, output and hidden layers for a $3$-dimensional communication signal is given in Fig. \ref{fig:e2emodel}.

\begin{figure*} [!t] 
\centering 
\includegraphics [width=6in]{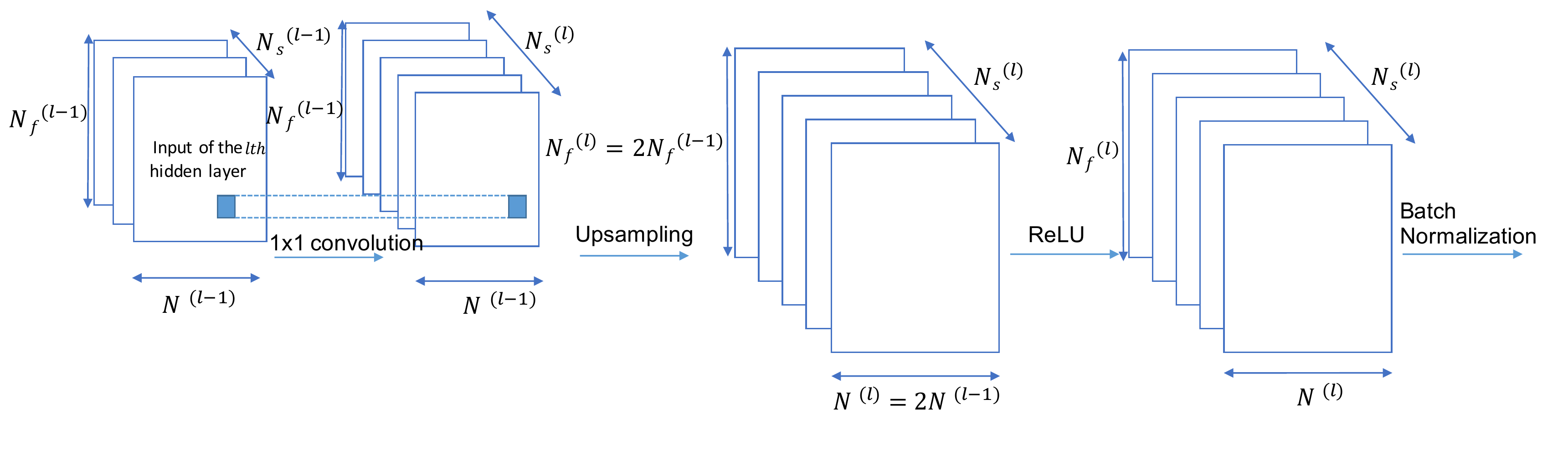}
\caption{The structure of the $l^{th}$ hidden layer, whose input dimension is $N_f^{(l-1)} \times N^{(l-1)} \times N_s^{(l-1)}$ and output dimension is $N_f^{(l)} \times N^{(l)} \times N_s^{(l)}$. Note that $N_f^{(l)} = 2N_f^{(l-1)}$ and $N^{(l)} = 2N^{(l-1)}$. The spatial dimensions $N_s^{(l-1)}$ and $N_s^{(l)}$ are the hyperparameters that are determined by the 1x1 convolution operations.} \label{fig:model}
\end{figure*}

The key component in the aforementioned DNN model is the hidden layers, which are composed of four major components. These are: (i) a 1x1 convolution, (ii) an upsampler, (iii) a rectified linear unit (ReLU) activation function, and (iv) a batch normalization. The 1x1 convolution means that each element in the time-frequency grid is processed with the same parameters through the spatial domain. That is, 1x1 convolution is used to change the dimension in spatial domain. This yields a tensor with dimension of $\mathbb{R}^{N_f^{(l)} \times N^{(l)} \times N_s^{(l)}}$ for the $l^{th}$ layer. In what follows, upsampling is performed to exploit the couplings among neighboring elements in the time and frequency grid. More precisely, the time-frequency signal is upsampled with a factor of $2$ via a bilinear transformation. Next, the ReLU activation function is used to make the model more expressive thanks to being a nonlinear function and allowing to pass the gradients over itself for low layers without saturating them. The last component of a hidden layer makes batch normalization for a batch size of $1$ to avoid the vanishing gradient problem. This structure of a hidden layer is portrayed in Fig \ref{fig:model}. All the hidden layers have the same structure except for the last hidden layer, which does not have an upsampler.

The mathematical representation of the aforementioned architecture is given next. Accordingly, the tensor $\mathbf{Y_T}$ is first parameterized as
\begin{equation}
\mathbf{\hat{Y}_T} = f_{\theta_l}(f_{\theta_{l-1}}( \cdots f_{\theta_0}(Z_0)))
\end{equation}
where the input $Z_0$ has a dimension of $N_f^{(0)} \times N^{(0)} \times N_s^{(0)}$ in the frequency, time and spatial domain, respectively. These are determined according to the number of hidden layers and the output dimension. The layers from $0$ to $l-1$ are counted as a hidden layer, and for $i=0,1,\cdots,l-2$
\begin{equation}
f_{\theta_{i}}=\text{BatchNorm}(\text{ReLU}(\text{Upsampler}(\theta_i\circledast Z_{i}))
\end{equation}
where $\circledast$ represents the convolution operator, which actually refers to cross-correlation in machine learning, and $Z_i$ is the input of the $i^{th}$ hidden layer. More precisely, a 1x1 convolution is utilized as a cross-correlator, which means that the spatial vector for each element of the time-frequency grid is multiplied with the same shared parameter matrix to obtain the new spatial vector for the next hidden layer. The last hidden layer is
\begin{equation}
f_{\theta_{l-1}}=\text{BatchNorm}(\text{ReLU}(\theta_{l-1}\circledast Z_{l-1}),
\end{equation}
and the output layer is
\begin{equation}
f_{\theta_{l}}=\theta_{l}\circledast Z_{l}.
\end{equation}

All the parameters can be represented as 
\begin{equation}
\Theta = (\theta_{0}, \theta_{1}, \cdots, \theta_{l}),
\end{equation}
which are optimized according to the square of $L2$-norm
\begin{equation}
\Theta^* = \arg \min_\Theta ||\mathbf{Y_T}-\mathbf{\hat{Y}_T}||_2^2.
\end{equation}
The output of the DNN for the optimized parameters is
\begin{equation}\label{outDNN}
\mathbf{Y_T^*} = f_{\Theta^*}(Z_0)
\end{equation}
where $ \mathbf{Y_T^*}= \{\{\mathbf{Y_T^*[m,n]}\}_{m=1}^{N_r}\}_{n=1}^{N}$. After generating the target signal in \eqref{outDNN} from a random input $Z_0$, an LS-type channel estimator is employed for the proposed deep channel estimator model using the pilots. Although our architecture is not restricted to any special pilot symbols arrangement, a block type pilot arrangement is utilized for brevity and concreteness. Accordingly, the pilots are sent at the beginning of each coherence time interval for all subcarriers. The channel can then be estimated between the $m^{th}$ receive antenna and the $q^{th}$ transmit antenna similar to the LS estimator given by
\begin{equation}\label{DCE}
\mathbf{\hat{H}_{DCE}[m,q]}= \mathbf{Y_T^*[m,1]}/\mathbf{X[q,1]}.
\end{equation}

\begin{figure*}[!t]
\centering
\subfigure[]{
\label{fig:threetaps}
\includegraphics[width=3.3in]{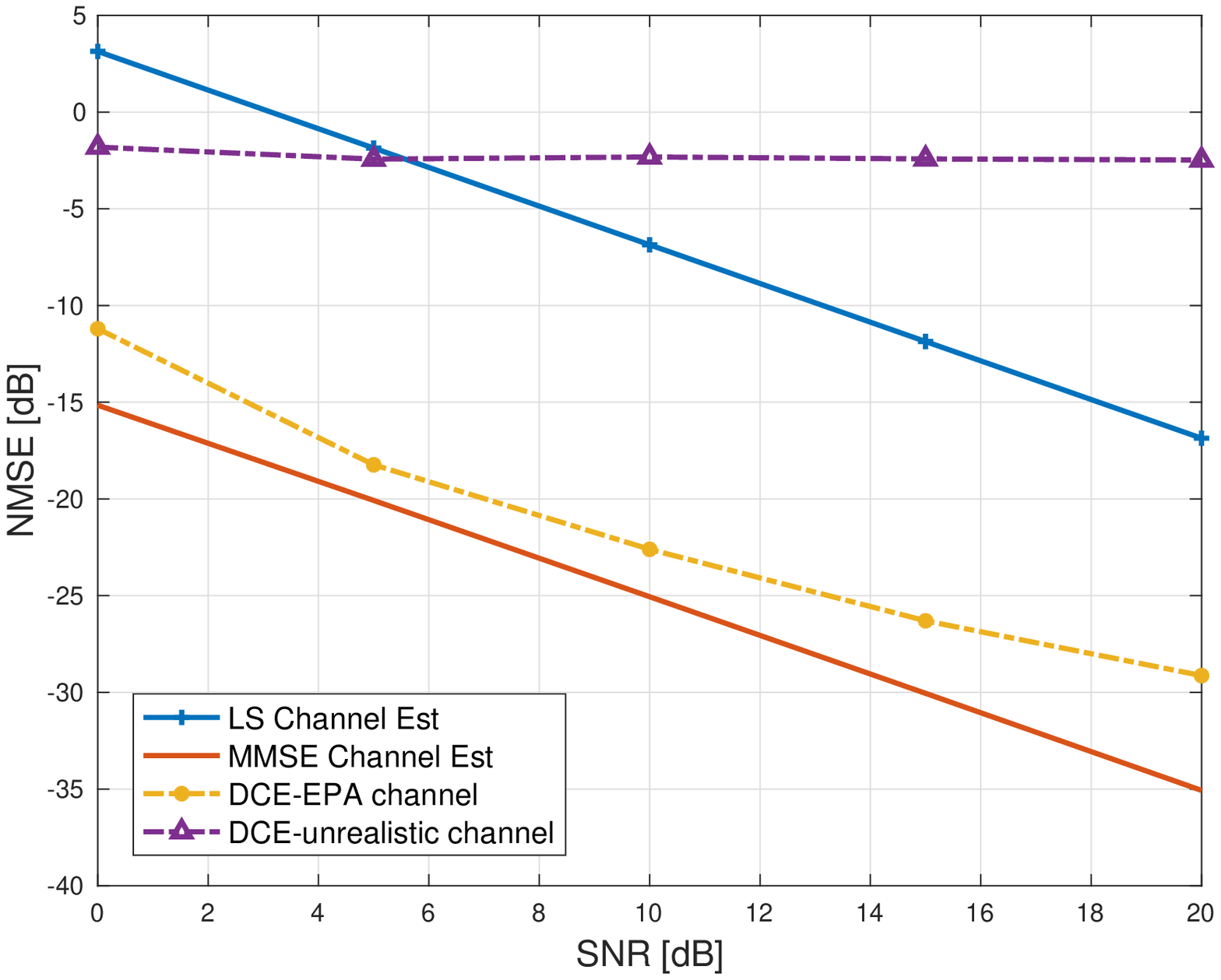}}
\qquad
\subfigure[]{
\label{fig:tentaps}
\includegraphics[width=3.3in]{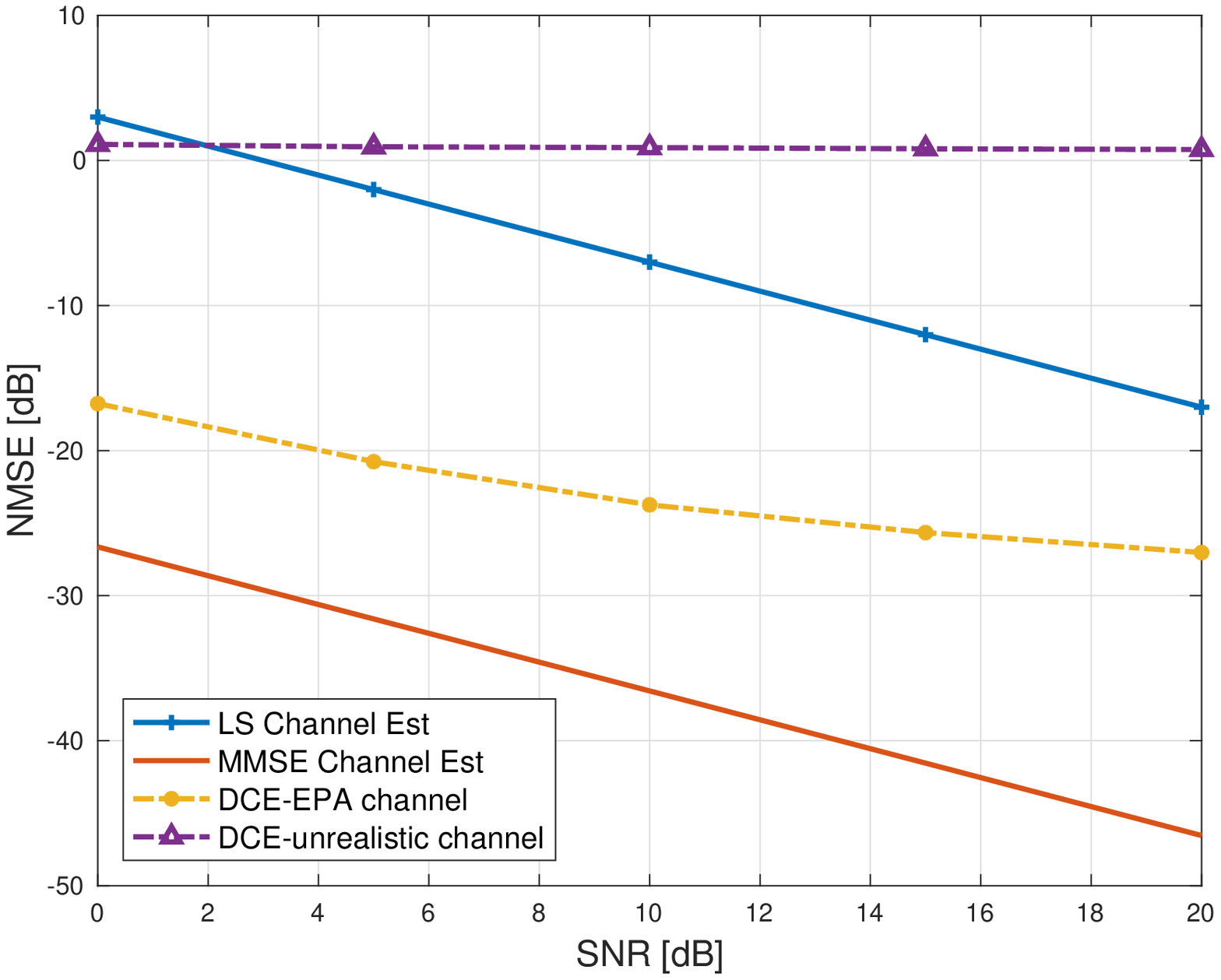}}
\caption{Channel estimation performance for SISO-OFDM with respect to SNR for (a) $64$ subcarriers and (b) $1024$ subcarriers.}\label{fig:SISO-OFDM}
\end{figure*}
\section{Simulations} \label{Simulations}
In this part, the proposed deep channel estimator in \eqref{DCE} is compared with the traditional LS and MMSE channel estimators given in \eqref{LS_est_vect} and \eqref{mmse_chn_est} using a realistic ``Extended Pedestrian A Model (EPA)'' channel. The performance metric is the normalized mean square error (NMSE), which is defined as 
\begin{equation}
\text{NMSE[m,q]} = \frac{||\mathbf{H[m,q]}-\mathbf{\hat{H}[m,q]}||^2}{||\mathbf{H[m,q]}||^2}
\end{equation}
where $\mathbf{H[m,q]}$ and $\mathbf{\hat{H}[m,q]}$ are the column vectors that specify the actual and the estimated channel taps in the frequency domain, respectively. Note that $\mathbf{\hat{H}[m,q]}$ can correspond to $\mathbf{\hat{H}_{LS}[m,q]}$, $\mathbf{\hat{H}_{MMSE}[m,q]}$ and $\mathbf{\hat{H}_{DCE}[m,q]}$. We consider two different scenarios. In the first one, a single antenna transmitter sends OFDM symbols to a single antenna receiver (SISO-OFDM) over the EPA channel model. In the second case, $K$ single antenna users transmit OFDM symbols to a base station that has $16$ antennas (massive MIMO-OFDM). The actual number of users is not important as long as $K\ll 16$ and they send orthogonal pilots. The simulations are done for $64$ and $1024$ subcarriers. 

The number of parameters of the proposed deep channel estimator is the same for $64$ and $1024$ subcarriers, which is possible thanks to the convolutional layer that utilizes the same parameters for all subcarriers. There are $6$ hidden layers in the proposed estimator. The input tensors are $2\times 2 \times128$ and $32\times32\times128$ for $64$ and $1024$ subcarriers because of oversampling with 2 in the time-frequency axes for the hidden layers except the last one. Note that the spatial dimension is determined arbitrarily, i.e., it is not optimized. The number of epochs for gradient descent is selected as $200$ for all cases to make fair comparisons.

\subsection{SISO-OFDM}
The efficiency of the deep channel estimator is depicted in Fig. \ref{fig:SISO-OFDM} for SISO-OFDM with $64$ and $1024$ subcarriers. As can be observed, our method outperforms LS channel estimation for all SNRs, and can approach the performance of MMSE channel estimation for EPA channel. It is worth pointing out that it would be more fair to compare the efficiency of the deep channel estimator with LS estimator by considering MMSE estimator as a lower bound. This is because MMSE estimator assumes that the second-order statistics of the channel and noise are known, which is not the case for the proposed and LS estimator. The proposed estimator gives up to $1$ dB close performance relative to the MMSE estimator in case of $64$ subcarriers. Although the gap increases for $1024$ subcarriers, this can be decreased by increasing the number of parameters in the DNN. 

\begin{figure} [!t] 
\centering 
\includegraphics [width=3.5in]{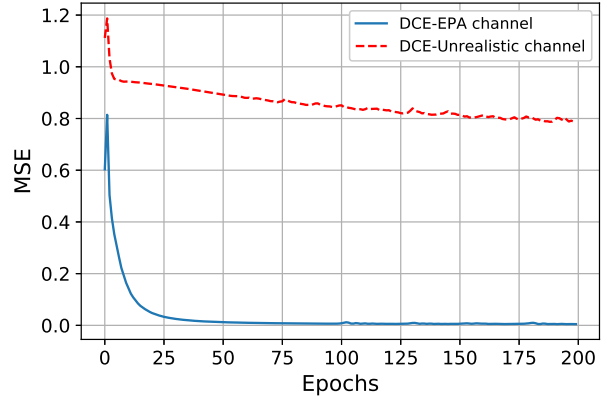}
\caption{The MSE behavior of the deep channel estimator for realistic EPA channel and unrealistic channel that has i.i.d. Gaussian taps in the frequency domain at $0$ dB.}\label{fig:DCE_MSE}
\end{figure}

To explain the underlying factor behind the superior performance of the deep channel estimator, we also tested it for an unrealistic channel model, which refers to the channel whose taps in the frequency domain are i.i.d. Gaussian random variables with zero mean and unit variance. This is unrealistic, because there has to be some correlations among the OFDM subcarriers. The proposed method does not give a satisfactory result in this case as can be seen in Fig. \ref{fig:SISO-OFDM}. This clearly demonstrates that the deep channel estimator exploits the correlations among subcarriers. To provide more insights, the decrease in the MSE with respect to the number of epochs while fitting the parameters is illustrated for both EPA and unrealistic channels in Fig. \ref{fig:DCE_MSE} at $0$ dB. There is a significant difference in MSE between these two channels, and in particular $200$ epochs are not sufficient when the channel taps in the frequency domain are independent. Note that if the number of epochs are increased, the deep channel estimator can perform slightly better with respect to LS estimator for all SNRs. This is associated with exploiting the correlations in the time domain.

\subsection{Massive MIMO-OFDM}
One of the salient features of the deep channel estimator reveals when it comes to reducing the pilot tones in LTE grid. More precisely, the deep channel estimator only needs $64$ pilots while processing an OFDM grid of $64\times64$ assuming the channel coherence time is greater than $4.5$ms. This reduces the number of pilots in the LTE grid by $68\%$, because $4$ pilots are allocated for each $84$ resource elements in LTE. This saving increases for larger coherence time such that our method can reduce the number of pilots $98\%$ in case of $1024$ subcarriers provided the channel coherence time is greater than $73$ms. This is particularly important for massive MIMO that lacks of sufficient pilots, which limits the performance. This can be leveraged both for the FDD and TDD modes in massive MIMO. To be more precise, the saving in the number of pilots can be equally shared among multiple antennas to send pilots in downlink channel estimation. This enables to employ the massive MIMO in FDD mode, and rules out some fundamental problems in TDD such as the need for the calibration of the hardware chains. Our method can also be used for the sake of TDD by enhancing the quality of the CSI in the uplink. 

\begin{figure} [!t] 
\centering 
\includegraphics [width=3.5in]{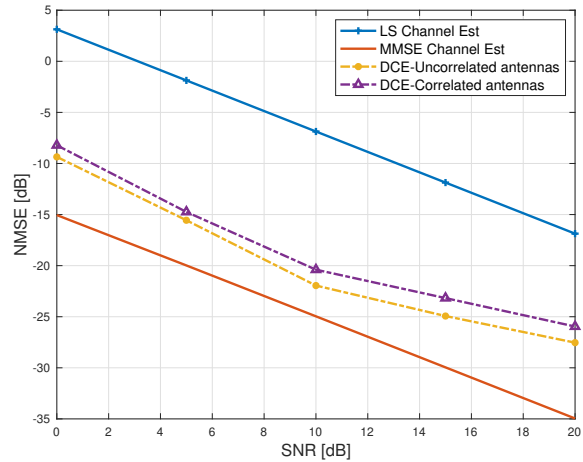}
\caption{Channel estimation performance for massive MIMO-OFDM with respect to SNR for $64$ subcarriers.}\label{fig:massive_MIMO_64}
\end{figure}

In this paper, the proposed method is utilized for the sake of an uplink massive MIMO channel estimation, in which $K$ users transmit to a $16$ antenna base station. For brevity, we only consider an OFDM symbol with $64$ subcarriers for this case, but it is straightforward to generalize this for higher number of subcarriers. The result is depicted in Fig. \ref{fig:massive_MIMO_64}. Similar to the SISO-OFDM, a very promising result was obtained to have near MMSE estimator performance with LS complexity. To highlight the impact of the antenna correlations, our result includes the cases when the receiver antennas are correlated and uncorrelated. For the former, an exponential correlation model is considered, in which the correlation coefficient is $\rho=0.5$ \cite{Loyka}. One important point is that the antenna correlations does not improve the performance of the deep channel estimator. This is because our architecture is designed to exploit the correlations in the time and frequency domain, i.e., oversampling is applied in the time-frequency grid.

\section{Conclusions} \label{Conclusions}
The deep channel estimator generates a less noisy signal from the received signal by fitting the parameters of the specially designed DNN. Then, this generated signal is element-wise divided to the pilot symbols so as to estimate the channel just like an LS estimation. Our results demonstrate that although the proposed channel estimator does not require the second-order statistics of the channel and noise, it can achieve within a $1$ dB  performance gap relative to MMSE estimation. The other intriguing features of the deep channel estimator lies in scalability in terms of the number of DNN parameters (e.g. independent of the number of subcarriers), great savings in the number of required pilot tones, and flexibility to be used in various channels with an arbitrary number of antennas. As future work, the benefit of the proposed deep estimator will be quantified for massive MIMO in terms of energy and spectral efficiency. It would also be interesting to observe and modify the deep channel estimator when there is pilot contamination for massive MIMO-OFDM or co-channel interference for SISO-OFDM.

\end{document}